\pdfoutput=1

\documentclass[11pt]{article}

\usepackage[preprint]{acl}

\usepackage{times}
\usepackage{latexsym}

\usepackage[T1]{fontenc}

\usepackage[utf8]{inputenc}

\usepackage{microtype}

\usepackage{inconsolata}

\usepackage{graphicx}

\usepackage{array} 
\usepackage{booktabs} 
\usepackage{multirow} 
\usepackage{colortbl} 
\usepackage{amsmath}  
\usepackage{subcaption}
\usepackage{amssymb}

\graphicspath{{./figs/}}

\title{A Semantic Search Engine for Mathlib4}

\author{
  \textbf{Guoxiong Gao\textsuperscript{1}}\thanks{These authors contributed equally.},
  \textbf{Haocheng Ju\textsuperscript{2}}\footnotemark[1],
  \textbf{Jiedong Jiang\textsuperscript{2}},
  \textbf{Zihan Qin\textsuperscript{1}},
  \textbf{Bin Dong,\textsuperscript{3,4}}\thanks{Corresponding author}
\\
\\
 \textsuperscript{1}School of Mathematical Sciences, Peking University, Beijing, China\\
 \textsuperscript{2}Beijing International Center for Mathematical Research, Peking University, Beijing, China\\
  \textsuperscript{3}Beijing International Center for Mathematical Research\\ and the New Cornerstone Science Laboratory, Peking University, Beijing, China\\
  \textsuperscript{4}Center for Machine Learning Research, Peking University, Beijing, China
\\
  \texttt{samggx@stu.pku.edu.cn} \quad
 \texttt{hcju@pku.edu.cn} \\
 \texttt{\{emailboxofjjd,15261375660\}@163.com} \quad
 \texttt{dongbin@math.pku.edu.cn}
}

\begin{document}

\maketitle

\begin{abstract}
The interactive theorem prover Lean enables the verification of formal mathematical proofs and is backed by an expanding community. Central to this ecosystem is its mathematical library, mathlib4, which lays the groundwork for the formalization of an expanding range of mathematical theories. However, searching for theorems in mathlib4 can be challenging. To successfully search in mathlib4, users often need to be familiar with its naming conventions or documentation strings. Therefore, creating a semantic search engine that can be used easily by individuals with varying familiarity with mathlib4 is very important. In this paper, we present a semantic search engine\footnote{The search engine is available at \href{https://leansearch.net}{https://leansearch.net}.} for mathlib4 that accepts informal queries and finds the relevant theorems. We also establish a benchmark for assessing the performance of various search engines for mathlib4.\footnote{The data and the source code are available at \href{https://github.com/reaslab/LeanSearch}{https://github.com/reaslab/LeanSearch}.}

\end{abstract}

\section{Introduction}

Lean \citep{DBLP:conf/cade/MouraKADR15,DBLP:conf/cade/Moura021} is an interactive theorem prover built on dependent type theory, designed to verify mathematical proofs written in a formal language and thereby enhancing their rigor. It has a vibrant and supportive community, with its popularity growing among mathematicians. A prime example of the Lean community's collaborative spirit is mathlib4, the mathematical library for Lean 4. This library, regularly updated by contributors from around the globe, acts as a basis for the formalization of new mathematical theories. This eliminates the need to repeatedly formalize established results, as users can simply check if mathlib4 contains the necessary theorems referenced in informal proofs. However, locating these theorems is often challenging due to the limitations of officially provided search tools, which struggle to find relevant theorems with informal queries.

There are primarily two methods to search for theorems in mathlib4: using the mathlib4 documentation\footnote{\href{https://leanprover-community.github.io/mathlib4_docs/}{https://leanprover-community.github.io/mathlib4\_docs/}} and searching the GitHub repository of mathlib4\footnote{\href{https://github.com/leanprover-community/mathlib4}{https://github.com/leanprover-community/mathlib4}}. The mathlib4 documentation allows users to search for theorems by their formal names. However, this feature can be challenging for beginners to utilize effectively, as they may not be familiar with the naming conventions. For example, Cauchy's Mean Value Theorem is named as \texttt{exists\_ratio\_deriv\_eq\_ratio\_slope} in mathlib4, meaning a direct search for "Cauchy's Mean Value Theorem" yields no results. An alternative method involves searching within the GitHub repository of mathlib4, which allows for keyword-based searches across the source files, including formal statements, proofs, and documentation strings. This method can locate Cauchy's Mean Value Theorem as it is mentioned in the documentation string of \texttt{exists\_ratio\_deriv\_eq\_ratio\_slope}. However, this approach faces two issues: many theorems in mathlib4 lack documentation strings, and semantically similar user queries that don't exactly match the theorems or documentation strings may lead to unsuccessful searches.

Consequently, neither method adequately supports finding theorems based on informal queries, leading to beginners spending significant time on this task. Discussions on Zulip\footnote{\href{https://leanprover.zulipchat.com/\#narrow/stream/219941-Machine-Learning-for-Theorem-Proving/topic/Semantic.20Search.20for.20Mathematics}{https://leanprover.zulipchat.com/\#narrow/stream/219941-Machine-Learning-for-Theorem-Proving/topic/Semantic.20Search.20for.20Mathematics}} have highlighted the need for creating a semantic search engine for mathlib4. Therefore, developing such a search engine for mathlib4 is highly desirable to improve the efficiency of theorem retrieval.

In this paper, we introduce a semantic search engine for mathlib4 that allows users to input an informal query and retrieve a list of relevant theorems from mathlib4. To construct this search engine, we translate the formal statements of mathlib4 theorems into informal ones and integrate these pairs into a database. Each database entry consists of a formal theorem statement and its informal version. Upon receiving a user query, we augment the query for improved context understanding and perform semantic search across the database to find relevant results. Additionally, we have established a benchmark to compare the effectiveness of various search engines for mathlib4.

The remaining part of the paper is organized as follows. In Section \ref{sec:related}, we review the related works on text retrieval and mathematical information retrieval. Section \ref{sec:method} describes our approach to developing a semantic search engine for mathlib4. Section \ref{sec:benchmark} presents the mathlib4 semantic search benchmark. Numerical results are discussed in Section \ref{sec:experiments}. We conclude this paper in Section \ref{sec:conclusion}.

\section{Related Work}\label{sec:related}

\subparagraph*{Text Retrieval.}
Text retrieval is the task of finding relevant information within a corpus based on user queries. Early methods, such as BM25 \citep{robertson1995okapi,10.1561/1500000019}, used sparse vector representations for queries and documents, assessing relevance by comparing these vectors with certain weighting techniques \citep{SALTON1988513}. While effective in measuring lexical similarity, these approaches fall short in capturing the semantic similarity between queries and documents.

To address this limitation, deep learning techniques \citep{10.1145/2505515.2505665,10.1145/3077136.3080809,mcdonald-etal-2018-deep,liu-etal-2018-entity} have been introduced. Utilizing deep neural networks, these techniques encode queries and documents into dense vectors, thereby assessing relevance based on semantic similarity. Further developments have been made in neural architectures and training paradigms. There are two main architectures: the cross-encoder \citep{DBLP:journals/corr/abs-1904-07531,nogueira-etal-2020-document} and the bi-encoder \citep{karpukhin-etal-2020-dense,qu-etal-2021-rocketqa,ni-etal-2022-large,ni-etal-2022-sentence}. Cross-encoders take the concatenation of query and document as input and produce the final relevance of this query-document pair, while bi-encoders map the query and document into vectors separately, determining relevance through similarity between the two vectors. Training paradigms have also evolved, with the Inverse Close Task (ICT) \citep{lee-etal-2019-latent} initially proposed for dense retriever pre-training. Subsequently, other pre-training tasks have been developed, including Body First Selection and Wiki Link Prediction, both introduced in \citep{DBLP:conf/iclr/ChangYCYK20}. Recent studies \citep{DBLP:journals/corr/abs-2201-10005,DBLP:journals/corr/abs-2212-03533,su-etal-2023-one,DBLP:journals/corr/abs-2309-07597} have explored large-scale unsupervised pre-training using contrastive loss, followed by fine-tuning on smaller, labeled datasets. Pre-training with extensive text pairs allows the language model to grasp textual semantics and the fine-tuning stage further enhances its performance across various retrieval tasks. However, the authors of \citep{DBLP:journals/corr/abs-2401-00368} argue that a two-stage training approach might not be necessary, demonstrating that directly fine-tuning a decoder-only model on both synthetic and labeled data can yield competitive results.

\subparagraph*{Mathematical Information Retrieval.}
Mathematical Information Retrieval (MIR) differs from text retrieval in that it involves queries and documents that contain mathematical formulas. These formulas are highly structured, which distinguishes them from plain text. To effectively capture the semantics of math formulas, several representations are employed. The Symbol Layout Tree (SLT) \citep{DBLP:journals/ijdar/ZanibbiB12} preserves the original layout of formulas, while the Operator Tree (OPT) \citep{DBLP:conf/ntcir/GaoYWJT16} represents mathematical symbols as nodes, with edges denoting the relationships between operators and operands. Classical MIR methods \citep{10.1145/3209280.3209527,DBLP:conf/ntcir/GaoYWJT16,DBLP:conf/ntcir/KristiantoTA16,10.1145/2911451.2911512,DBLP:conf/clef/KaneNT22,DBLP:conf/clef/NgFKT21,DBLP:conf/clef/NgFKLMTW20} rely on structure search, identifying matching substructures across various features. Among these, the Approach0 structural search method \citep{DBLP:conf/ecir/ZhongZ19,10.1007/978-3-030-45439-5_47} has shown to be particularly effective. It indexes formulas through leaf-root paths in the OPT and utilizes subexpression matching to assess formula similarity. With advancements in deep learning, dense retrievers have been integrated into MIR, often combined with structural searches \citep{DBLP:conf/clef/KaneNT22,DBLP:conf/clef/Zhong0L22,DBLP:conf/emnlp/ZhongY0L22,10.1145/3539618.3591746}. A notable example is the Approach0 hybrid search \citep{DBLP:conf/clef/Zhong0L22}, which combines Approach0 structure search with a bi-encoder dense retriever, ColBERT \citep{10.1145/3397271.3401075}. This combination not only facilitates effective formula matching but also enhances understanding of context.

The preceding literature focuses on extracting mathematical content from a corpus composed of natural language texts and formulas formatted in markup languages. This contrasts with our objective of conducting searches within mathlib4, a formal mathematical library. The work most closely related to ours is Moogle\footnote{\href{https://www.moogle.ai/}{https://www.moogle.ai/}}, a semantic search engine for mathlib4. However, its technical details have not been released. We will compare the performance of Moogle and our search engine in Section \ref{sec:experiments}.

\section{Methodology}\label{sec:method}
\begin{figure}[t]
\centering
\includegraphics[width=\columnwidth]{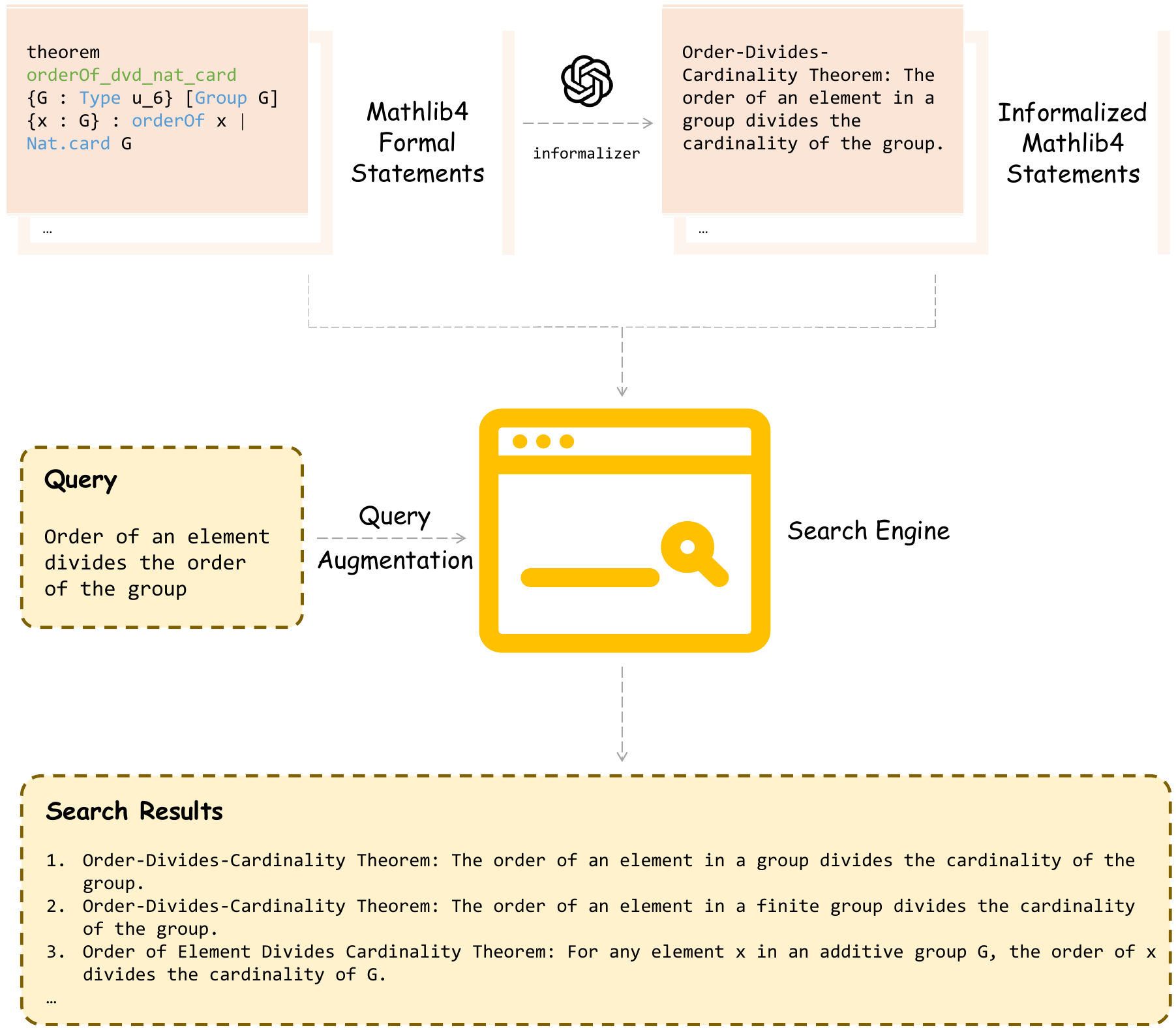}
\caption{Overview of our method for creating a semantic search engine for mathlib4. We employ an informalizer to convert formal statements from mathlib4 into their informal counterparts. These informal-formal pairs are then stored in a vector database. When users input a query, the system augments the query and searches across the database, yielding a list of relevant theorems.}
\label{fig:overview}
\end{figure}

In this section, we will describe the implementation of our semantic search engine for mathlib4\footnote{We use mathlib4 with commit \texttt{db04a978b67b2200691f0bd968f334c83261b66a}.}. This engine is designed to accept a user query in natural language and return a list of relevant theorems in mathlib4. Our approach involves converting formal theorems from mathlib4 into their informal counterparts, as illustrated in Figure \ref{fig:overview}. These informal-formal theorem pairs are then vectorized and stored in a vector database, a step that can be executed offline. When a user query is submitted, we augment the query to better grasp its context, vectorize the enriched query, and locate the corresponding theorems in the embedding space. In the following subsections, we will elaborate on the informalization of mathlib4, the design of the search engine, and the method of query augmentation.

\subsection{Informalizing Mathlib4}
\begin{figure}[t]
\centering
\includegraphics[width=\columnwidth]{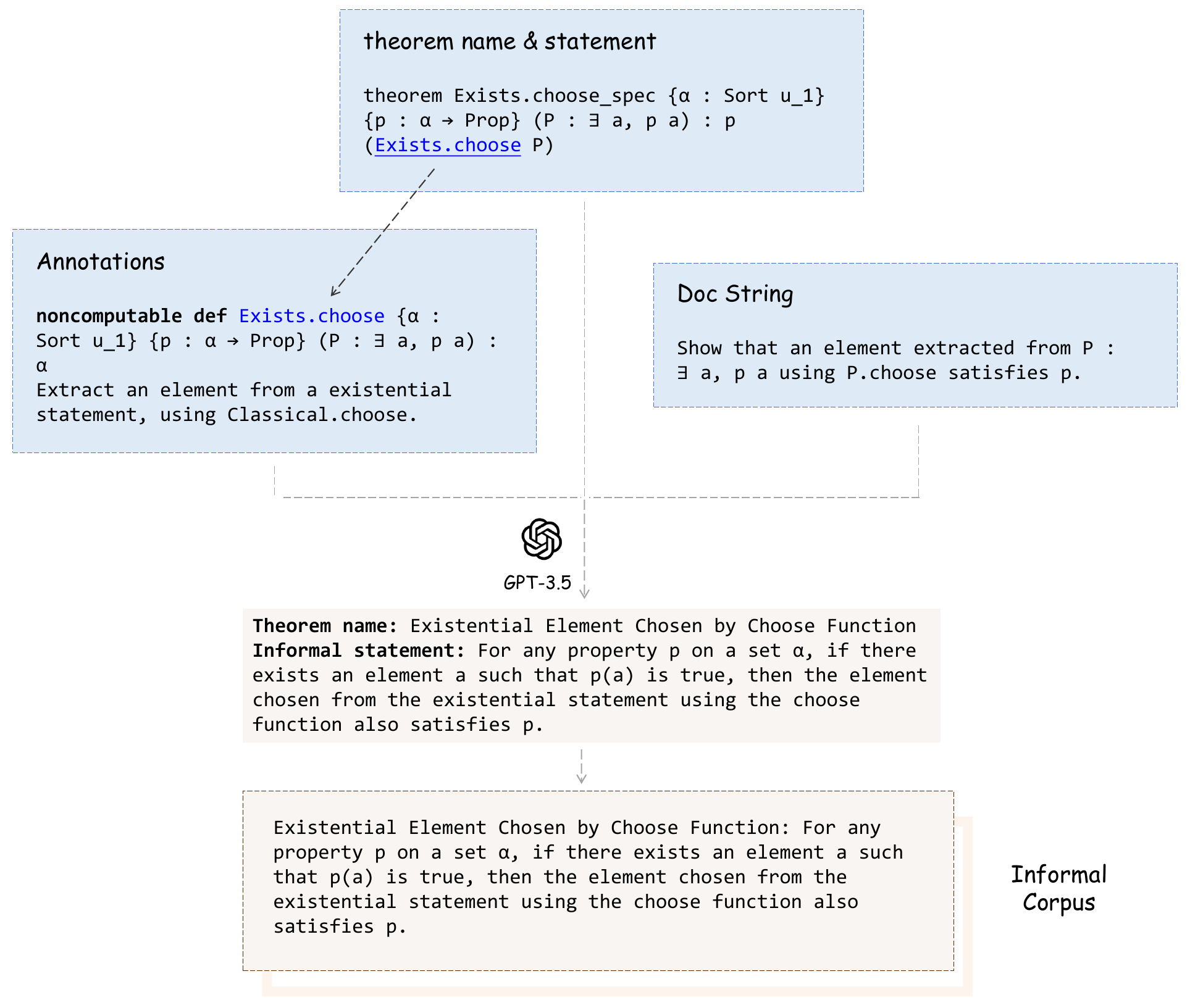}
\caption{Our approach to informalizing mathlib4 theorems. We extract the theorem name, statement, and documentation string from the mathlib4 documentation. Moreover, we collect related definitions via the hyperlinks in the theorem statements. The gathered information is then inputted into GPT-3.5 to generate informal names and statements. These are then organized in the format "theorem name: informal statement" and added to an informal corpus.}
\label{fig:informalize}
\end{figure}

\begin{figure}[t]
\centering
\includegraphics[width=\columnwidth]{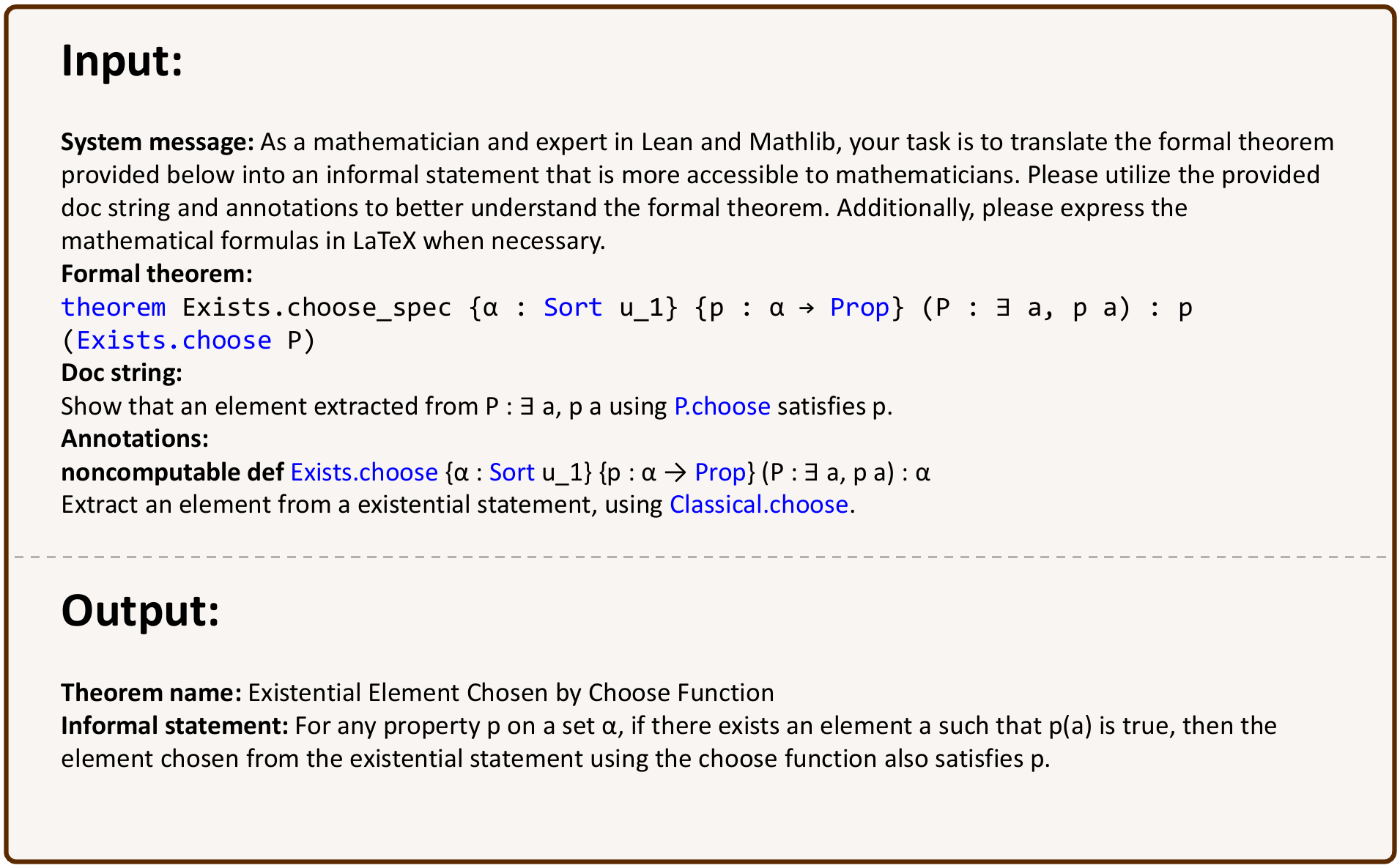}
\caption{Prompt for informalizing mathlib4 statements with documentation strings.}
\label{fig:prompt_ex1}
\end{figure}

Our strategy for informalizing mathlib4 involves employing a large language model (LLM). Central to this strategy is providing the LLM with sufficient context to accurately grasp the formal theorem's exact meaning. To this end, we not only extract the theorem's name, statement, and documentation string\footnote{For many theorems in mathlib4 that lack a documentation string, we will extract only the theorem name and statement.} from the mathlib4 documentation but also include related definitions found in the theorem statements through hyperlinks. For example, as illustrated in Figure \ref{fig:informalize}, we extract the definition of \texttt{Exists.choose} because it is referenced and linked in the theorem \texttt{Exists.choose\_spec}. The gathered information is then fed into an LLM\footnote{We use \texttt{gpt-3.5-turbo-16k} for generating informal names and statements, setting the temperature to $0$.} to generate informal theorem names and statements. These are subsequently formatted as "theorem name: informal statement" and incorporated into an informal corpus. Figure \ref{fig:prompt_ex1} shows the prompt used for informalizing the theorem \texttt{Exists.choose\_spec}, including the formal statement, documentation string, and the definition of \texttt{Exists.choose} as annotations to aid the LLM's understanding.

Notably, the authors of \citep{DBLP:journals/corr/abs-2311-03755} also employed an LLM for informalizing mathlib4 statements, relying solely on the formal statements. We argue that providing the LLM with additional context, such as related definitions and documentation strings, enhances its ability to accurately interpret and convert formal theorems into their informal counterparts. An example comparing the informalized mathlib4 statement generated by \citep{DBLP:journals/corr/abs-2311-03755} to the one produced by our method is provided in Appendix \ref{sec:example}.

\subsection{Semantic Search Engine for Mathlib4}
After obtaining the informal corpus, we employ dense embedding models, which excel at capturing semantic information, to encode the informal-formal theorem pairs. Recent advancements in text embedding models have introduced the practice of integrating specific task instructions into either queries or documents, enhancing the model's adaptability to diverse tasks and boosting performance in zero-shot settings \citep{su-etal-2023-one,DBLP:journals/corr/abs-2309-12871,DBLP:journals/corr/abs-2401-00368}. Consequently, for the purpose of theorem retrieval within mathlib4, we enrich our corpus documents with specific instructions. We adopt the following instruction template:
\begin{center}
    \texttt{Instruct: Retrieve math theorems stated in bilingual Lean 4 + natural language that are mathematically equivalent to the given one \textbackslash n Doc:\{document\}}
\end{center}
Here, \texttt{\{document\}} denotes

\begin{center}
    \texttt{"\{formal statement\} \textbackslash n \{informal name\}:\{informal statement\}"}.
\end{center}

These task-specific instructions significantly influence the performance of embedding models. We will investigate the effect of varying task instructions on the overall retrieval performance in Appendix \ref{sec:append_abl}.

Embedding the entire corpus, although a time-consuming process, is performed offline and does not require repetition for each use of the engine. We utilize Chroma DB to store the embeddings. Upon receiving a query, our system vectorizes it and retrieves theorems based on their cosine distance to the query in the embedding space. Chroma DB employs the Hierarchical Navigable Small World (HNSW) algorithm \citep{DBLP:journals/pami/MalkovY20}, an efficient approximate nearest neighbor search method, ensuring rapid retrieval from the corpus.

\subsection{Query Augmentation}

To enhance query clarity and achieve a more accurate embedding, our system incorporates a query augmentation step. This process involves prompting an LLM\footnote{We use \texttt{gpt-4-0125-preview} for query augmentation.} to transform a brief, vague query into a detailed statement that includes both informal and formal statements, ensuring mathematical equivalence with the original query. We guide the LLM with specific principles for query augmentation, emphasizing the importance of precision, the use of LaTeX for mathematical expressions, and the clarification of ambiguous inputs. Additionally, we provide examples of query augmentations to improve the LLM's comprehension of the task. A detailed prompt for query augmentation is provided in Appendix \ref{sec:append_qa_prompt}. Although Lean 4 code generation by LLMs may occasionally introduce inaccuracies due to the limited presence of mathlib4 in their training data, this approach effectively enriches the query with additional contextual information. 

Following augmentation, the enriched query is structured as \texttt{"\{formal statement\} \textbackslash n \{informal name\}:\{informal statement\}"}, matching the structure of our database. In a similar manner to adding task instructions in document processing, we enrich the query with specific instructions, utilizing a template as follows:
\begin{center}
    \texttt{Instruct: Retrieve math theorems stated in bilingual Lean 4 + natural language that are mathematically equivalent to the given one \textbackslash n Query:\{formal statement\} \textbackslash n \{informal name\}:\{informal statement\}"}
\end{center}
This formatted query is then vectorized and utilized by the search engine to retrieve theorems based on the cosine distance in the embedding space.

\section{Mathlib4 Semantic Search Benchmark}\label{sec:benchmark}

To rigorously assess and compare the efficacy of various retrieval methods, we have established the Mathlib4 Semantic Search Benchmark. This benchmark encompasses a curated set of queries, relevance labels of each mathlib4 theorem for these queries, and a collection of performance metrics. In this section, we provide a detailed explanation of the benchmark. Section \ref{subsec:query} describes the composition of the query set. In Section \ref{subsec:relevance}, we explain the relevance criteria for our benchmark and the labeling procedure. Section \ref{subsec:metrics} lists the performance metrics used in our benchmark.

\subsection{Composition of the Query Set}\label{subsec:query}

The query set of our benchmark has 50 distinct queries, spanning various mathematical disciplines including calculus, abstract algebra, linear algebra, number theory, algebraic number theory, set theory, and mathematical logic. This selection aims to cover a broad spectrum of topics and complexities. We argue that this is a typical size for the query sets in mathematical information retrieval datasets, as demonstrated by the ARQMath1, ARQMath2, and ARQMath3 databases from the MIR field, which contain 77, 71, and 78 queries, respectively, for their answer retrieval tasks \citep{Mansouri2022OverviewOA}.

To optimize labeling efforts, we organize queries with identical search intents into 18 distinct, non-overlapping groups, each containing at least two distinct queries.
This approach assumes that all queries within a group have the same relevance score for any document, significantly reducing the need for repetitive labeling, as each document is evaluated only once per query group.
Furthermore, to provide a more detailed assessment and mitigate the impact of duplicate document labels, we consider four prevalent forms of mathematical queries: natural language descriptions, LaTeX formulas, theorem names, and Lean 4 term descriptions. In each query group, we strive to include as many different description forms as possible. Table \ref{table:Queries} provides statistics and examples of the query set. However, not all query groups contain all four representation forms, as some theorems may not have an official name or their representation in a LaTeX formula might be redundant.

\begin{table}[h!]
\centering
\resizebox{\linewidth}{!}{
\begin{tabular}
{
  |>{\centering\arraybackslash}m{1.8cm} 
  |>{\centering\arraybackslash}m{1.1cm} 
  ||>{\raggedright\arraybackslash}m{5.5cm}
  |>{\raggedright\arraybackslash}m{3.5cm}|
}
\hline
\textbf{Category} & \textbf{Count} & \textbf{Example 1} & \textbf{Example 2} \\ \hline
Natural Description & 18 & If there exist injective maps of sets from \(A\) to \(B\) and from \(B\) to \(A\), then there exists a bijective map between \(A\) and \(B\). & If \(p\) implies \(q\), then not \(q\) implies not \(p\). \\ \hline
LaTeX Formula & 15 & If there exist \(f : A \rightarrow B\) injective, \(g : B \rightarrow A\) injective, then there exists \(h : A \rightarrow B\) bijective. & \((p \rightarrow q) \rightarrow (\neg q \rightarrow \neg p)\) \\ \hline
Theorem Name & 7 & Schroeder Bernstein Theorem & Modus Tollens \\ \hline
Lean 4 Term & 10 & \texttt{\{f : A → B\} \{g : B → A\} (hf : Injective f) (hg : Injective g) : $\exists$ h, Bijective h} & \texttt{(p → q) → (¬q → ¬p)} \\ \hline
\end{tabular}
}
\caption{Statistics and examples of query groups in our benchmark.}
\label{table:Queries}
\end{table}

\subsection{Relevance Judgments}\label{subsec:relevance}
During the process of collecting relevance labels for query-document pairs, we adopt an approach similar to ARQMath\citep{Mansouri2022OverviewOA}. Initially, we establish a carefully crafted relevance assessment criteria, elaborated in Table \ref{table:Relevant_criteria}. While performing the labeling, assessors are instructed to evaluate whether a given search result facilitates their mathematical formalization workflow. Instead of merely considering similarity in topic, presented form, formula structure, or mathematical induction relationship, the relevance we focus on here is deeply aligned with the needs of Lean 4 experts.

\begin{table}[h!]
\centering
\resizebox{\linewidth}{!}{
\begin{tabular}{
  |>{\centering\arraybackslash}m{2cm}
  |>{\centering\arraybackslash}m{1cm}
  |>{\centering\arraybackslash}m{1cm}
  |>{\raggedright\arraybackslash}m{8cm}|
}
\hline
\textbf{Rating} & \textbf{Label} & \textbf{Score} & \textbf{Definition}\\ \hline
Exact Match & 2 & 1 & Exact match to the query or being a stronger statement \\ \hline
Relevant & 1 & 0.3 & Useful in locating where the corresponding statement should be \\ \hline
Irrelevant & 0 & 0 & Not expected to be useful in mathematics formalization workflow \\ \hline
\end{tabular}
}
\caption{The relevance assessment criteria for our benchmark.}
\label{table:Relevant_criteria}
\end{table}

For each set of queries grouped by identical search intentions, assessors are presented with the top 50 theorems retrieved by an intermediate version of our search engine. In addition to evaluating the provided theorems, assessors are tasked with identifying and adding any relevant theorems that may have been omitted from the initial list by inspecting the files where "Exact Match" items in the list are located, with particular attention to those with "Exact Match" rating. 

Given the structured organization of mathlib4, where related theorems are often located within the same file, it is reasonable to assume that any items not in the list are irrelevant to this query group. Additionally, all queries are guaranteed to have at least one exact match in mathlib4. These assumptions support the performance metrics we use in the benchmark, as described in the following subsection.

\subsection{Performance Measures}\label{subsec:metrics}

To compare different retrieval paradigms' performance on our labeled dataset, three commonly used metrics are adopted in our benchmark. Precision@k calculates the average relevant document proportion in top-$k$ retrieved results among the query set $\mathcal{Q}$:
\begin{equation}
\text{Precision@k} = \frac{1}{k|\mathcal{Q}|} \sum_{i=1}^{|\mathcal{Q}|} \sum_{j=1}^{k} \mathbb{I}(i,d_i^j),
\end{equation}
where $d_i$ represents the list of retrieved theorems for the $i$-th query, and $\mathbb{I}(i,d_i^j) = 1$ if and only if the $j$-th retrieved result of $i$-th query is "Exact Match", otherwise $\mathbb{I}(i,d_i^j) = 0$. Assuming that all unlabeled theorems are irrelevant, we can also calculate Recall@k:
\begin{equation}
\text{Recall@k} = \frac{1}{|\mathcal{Q}|} \sum_{i=1}^{|\mathcal{Q}|} \frac{1}{\sigma_i}\sum_{j=1}^{k} \mathbb{I}(i,d_i^j),
\end{equation}

where $\sigma_i$ is the number of "Exact Match" theorems for i-th query.

The third metric, nDCG (normalized Discounted Cumulative Gain), incorporates the position of the retrieved result. A decaying weight is allocated to each position:
\begin{equation}
\text{DCG}_i\text{@k} = \sum_{j=1}^{k} \frac{s(i,d_i^j)}{\log_2(j+1)},
\end{equation}

where $s(i,d_i^j)$ represents the score of the $j$-th retrieved theorem for the $i$-th query, as defined in Table \ref{table:Relevant_criteria}. Our scoring system is not proportional to the label number, emphasizing the importance of exact matching in the mathlib4 retrieval task. We further define $\text{IDCG}_i\text{@k}$ as the highest possible $\text{DCG}_i\text{@k}$ by properly arranging the retrieved theorems, and finally:
\begin{equation}
\text{nDCG}\text{@k} = \frac{1}{|\mathcal{Q}|}\sum_{i=1}^{|\mathcal{Q}|}\frac{\text{DCG}_i\text{@k}}{\text{IDCG}_i\text{@k}}
\end{equation}

To sum up, nDCG@k measures both the retrieval and ranking ability of the given engine in a unified and detailed way. Meanwhile, P@10 (Precision@10) and R@10 (Recall@10) focus solely on the retrieval effectiveness. In our benchmark, all three metrics will be reported for both the entire query set and each query category.

\section{Experiments}\label{sec:experiments}

In this section, we evaluate the performance of various theorem retrieval methods using our benchmark. The experimental setup is detailed in Section \ref{subsec:exp_setup}. We compare the performances of different retrieval methods in Section \ref{subsec:main_results}. Additionally, the ablation study on document preparations and query augmentations is presented in Section \ref{subsec:ablation}.

\subsection{Experiment Setup}\label{subsec:exp_setup}

We have considered four embedding models in our experiments: text-embedding-ada-002 and text-embedding-3-large from OpenAI, UAE-Large-V1\citep{DBLP:journals/corr/abs-2309-12871}, and $\text{E5}_{\text{mistral-7b}}$\citep{DBLP:journals/corr/abs-2401-00368}. For the baseline methods, we compare BM25 \citep{10.1561/1500000019}, Moogle, and the four embedding models applied to the original Lean 4 formal corpus and unaugmented query. The same four embedding models equipped with formal + informal query augmentation and formal + informal document corpus are also tested on our benchmark to demonstrate the efficacy of our approach. The implementation details are provided in Appendix \ref{sec:append_detail}.

\subsection{Main Results}\label{subsec:main_results}

Table \ref{table:main_results} and \ref{table:by_category} present the results of BM25, Moogle, and four embedding models. The model $\text{E5}_{\text{mistral-7b}}$, when integrated with our retrieval pipeline, achieves the best performance across three overall metrics and significantly outperforms all other methods, including its own performance on a formal corpus with unaugmented queries. We observe that all retrieval methods benefit from using an augmented corpus and queries, as augmentation expands concrete mathematics and Lean 4 terms, and these embedding models typically perform better on symmetrical retrieval tasks \citep{DBLP:journals/corr/abs-2401-00368}.

Examining the results averaged by category, we find that $\text{E5}_{\text{mistral-7b}}$ attains the best results in nearly all categories, with the exception of nDCG@20 for Lean 4 Terms. This exception is likely due to the augmented informal information negatively impacting retrieval performance, as it is not essential in formal-formal retrieval (formal queries, formal results) which focuses on lexical matches. Notably, our method significantly improves performance in the Theorem Name category. The enhancement is due to the expansion of the statement for a given theorem name by our query augmentation, which is essential for successful retrieval in this category. Furthermore, the combined corpus of formal and informal statements facilitates the retrieval process.

\begin{table}[h!]
\centering
\resizebox{\linewidth}{!}{
\begin{tabular}{lccccc}
\toprule
\textbf{Model} & \textbf{Corpus} & \textbf{Query aug.} & \textbf{nDCG@20} & \textbf{P@10} & \textbf{R@10}\\
\midrule
\multicolumn{6}{c}{\textbf{Baselines}}\\
\midrule
\textbf{Moogle\textsuperscript{\dag}} & \textbackslash & \textbackslash & 0.365\textsuperscript{\dag} & 0.092\textsuperscript{\dag} & 0.513\textsuperscript{\dag}\\
\textbf{BM25} & F & \textbackslash & 0.024 & 0.004 & 0.030\\
\textbf{OpenAI v2} & F & \textbackslash & 0.312 & 0.078 & 0.405\\
\textbf{OpenAI v3} & F & \textbackslash & 0.493 & 0.128 & 0.622\\
\textbf{UAE-Large-V1} & F & \textbackslash & 0.233 & 0.066 & 0.307\\
\textbf{$\text{E5}_{\text{mistral-7b}}$} & F & \textbackslash & 0.593 & 0.132 & 0.687\\
\midrule
\multicolumn{6}{c}{\textbf{Our Methods}}\\
\midrule
\textbf{OpenAI v2} & F+IF & F+IF & 0.553 & 0.144 & 0.707\\
\textbf{OpenAI v3} & F+IF & F+IF & 0.691 & 0.178 & 0.837\\
\textbf{UAE-Large-V1} & F+IF & F+IF & 0.368 & 0.084 & 0.440\\
\textbf{$\text{E5}_{\text{mistral-7b}}$} & F+IF & F+IF & \textbf{0.733} & \textbf{0.196} & \textbf{0.913}\\
\bottomrule
\end{tabular}
}
\caption{Results on our benchmark, averaged across all queries in our dataset. Here "aug.", F, IF, P@10 and R@10 represent augmentation, formal, informal, Precision@10, and Recall@10, respectively. The terms OpenAI v2 and v3 refer to the text-embedding models ada-002 and 3-large, respectively.  Moogle\textsuperscript{\dag}, unlike other retrieval systems, not only fetches theorems but also definitions, structures, instances, etc., making it incomparable under our performance metrics directly. For the purpose of this analysis, all non-theorem items retrieved are considered irrelevant, given the explicit theorem-searching intent of our queries. This approach, however, might advantage our theorem retrieval systems over Moogle, as non-theorem items occupy potential slots in the retrieval list. The notation \textsuperscript{\dag} is used to denote this adjustment.}
\label{table:main_results}
\end{table}

\begin{table}[h!]
\centering
\resizebox{\linewidth}{!}{
\begin{tabular}{lcccccccc}
\toprule
\multirow{2}{*}{\textbf{Model}} & \multicolumn{4}{c}{\textbf{nDCG@20}} & \multicolumn{4}{c}{\textbf{P@10}}\\
\cmidrule(r){2-5} \cmidrule(lr){6-9}
& ND & LF & TN & LT & ND & LF & TN & LT\\
\midrule
\multicolumn{9}{c}{\textbf{Baselines}}\\
\midrule
\textbf{Moogle\textsuperscript{\dag}} & 0.369\textsuperscript{\dag} & 0.324\textsuperscript{\dag} & 0.333\textsuperscript{\dag} & 0.441\textsuperscript{\dag} & 0.083\textsuperscript{\dag} & 0.107\textsuperscript{\dag} & 0.071\textsuperscript{\dag} & 0.100\textsuperscript{\dag} \\
\textbf{BM25} & 0.000 & 0.000 & 0.000 & 0.119 & 0.000 & 0.000 & 0.000 & 0.020\\
\textbf{OpenAI v2} & 0.276 & 0.379 & 0.000 & 0.498 & 0.061 & 0.107 & 0.000 & 0.120\\
\textbf{OpenAI v3} & 0.479 & 0.553 & 0.235 & 0.610 & 0.122 & 0.160 & 0.057 & 0.140\\
\textbf{UAE-Large-V1} & 0.301 & 0.216 & 0.004 & 0.298 & 0.078 & 0.067 & 0.000 & 0.090\\
\textbf{$\text{E5}_{\text{mistral-7b}}$} & 0.576 & 0.633 & 0.294 & \textbf{0.774} & 0.139 & 0.140 & 0.043 & 0.170\\
\midrule
\multicolumn{9}{c}{\textbf{Our Methods}}\\
\midrule
\textbf{OpenAI v2} & 0.536 & 0.533 & 0.571 & 0.600 & 0.111 & 0.160 & 0.186 & 0.150\\
\textbf{OpenAI v3} & 0.681 & 0.657 & 0.772 & 0.703 & 0.167 & 0.180 & 0.200 & \textbf{0.180}\\
\textbf{UAE-Large-V1} & 0.415 & 0.337 & 0.371 & 0.329 & 0.100 & 0.080 & 0.071 & 0.070\\
\textbf{$\text{E5}_{\text{mistral-7b}}$} & \textbf{0.748} & \textbf{0.712} & \textbf{0.855} & 0.654 & \textbf{0.194} & \textbf{0.200} & \textbf{0.214} & \textbf{0.180}\\
\bottomrule
\end{tabular}
}
\caption{Results on our benchmark averaged by category. Here ND, LF, TN and LT stands for Natural Description, LaTeX Formula, Theorem Name and Lean 4 Term respectively. Corpus type and query augmentation type remain the same as Table \ref{table:main_results} and are omitted here due to space limitations. \textsuperscript{\dag} indicates a different performance calculation rule, detailed in captions of Table \ref{table:main_results}.}
\label{table:by_category}
\end{table}

\subsection{Ablation Studies}\label{subsec:ablation}

In this subsection, we perform ablation studies on document preparation and query augmentation. The results of these studies are presented in Sections \ref{subsubsec:ablate_doc} and \ref{subsubsec:ablate_query}, respectively. Additionally, the ablation study on task instructions in provided in Appendix \ref{sec:append_abl}.

\subsubsection{Ablation of Document Preparations}\label{subsubsec:ablate_doc}

Table \ref{table:doc_query_ablation} presents the effects of changing corpus type with and without query augmentation. We use the original Lean 4 statement and its informalized version to create the formal and informal corpus. Without query augmentation, results indicate that both OpenAI v3 large and $\text{E5}_{\text{mistral-7b}}$ underperform with incomplete corpus components, yielding lower overall scores on all metrics. Specifically, the formal + informal corpus enhances performance on Lean 4 Terms by incorporating formal data and improves score on the Natural Description category by providing hybrid information. For the Theorem Name category, $\text{E5}_{\text{mistral-7b}}$ significantly benefits from the formal + informal corpus by utilizing the unfolded mathematical descriptions provided by the informalized statement, whereas OpenAI v3 large shows diminished results, likely due to the lack of training on mixed-domain texts and absence of adaptive instructions. We note that the formal corpus outperforms the formal+informal corpus in the Lean 4 Terms category for $\text{E5}_{\text{mistral-7b}}$, as explained in Section \ref{subsec:main_results}: retrieval on Lean 4 Terms relies heavily on lexical rather than semantic information, and the presence of informal information negatively impacts retrieval performance. Similar results are observed with augmented queries.

\begin{figure}
\centering
    \includegraphics[width=\columnwidth]{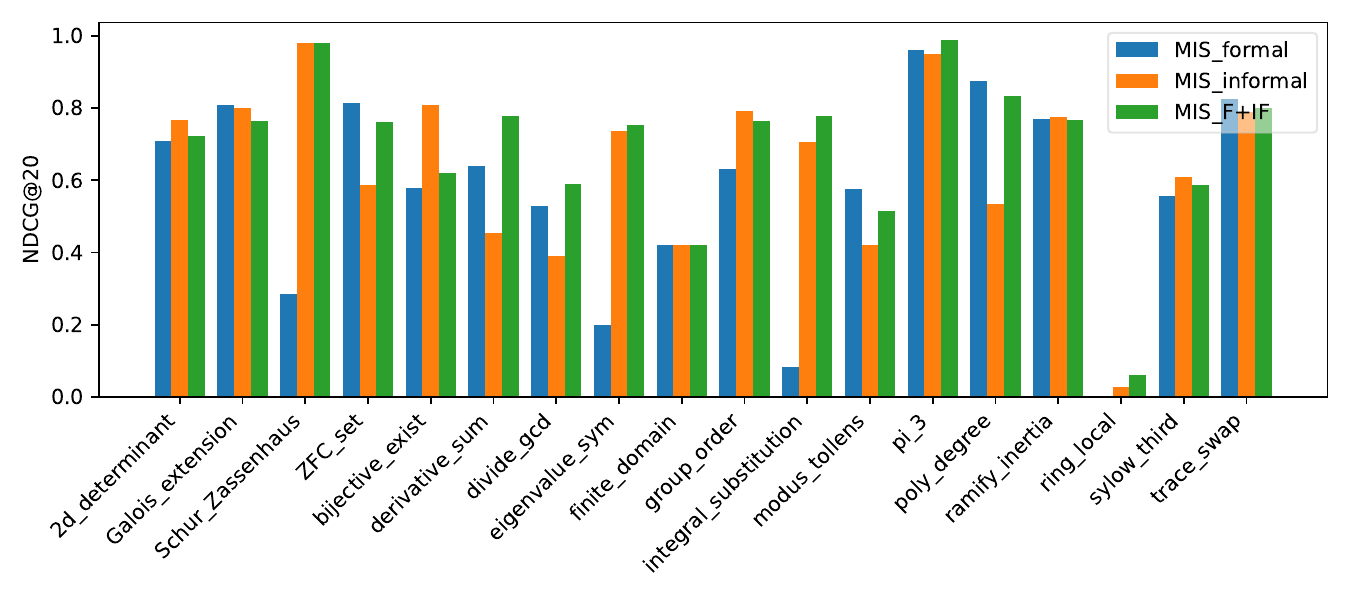}
    \caption{nDCG@20 performance of $\text{E5}_{\text{mistral-7b}}$ across formal, informal and formal + informal corpus on our benchmark, averaged by query groups. These evaluations are conducted using non-augmented queries.}
    \label{fig:doc_prep}
\end{figure}

A more detailed examination of the results for each query group without query augmentation is shown in Figure \ref{fig:doc_prep}. In most instances, the "formal + informal" configuration avoids the abrupt performance decline observed in solely formal or informal settings. The informalized statement provided in the corpus mitigates the lexical mismatch between the query and the formal statement, which justifies our choice of a combined corpus.

\begin{table}[h!]
\centering
\resizebox{\linewidth}{!}{
\begin{tabular}{lccccccc}
\toprule
\multirow{2}{*}{\textbf{Model \& Corpus}} & \multicolumn{5}{c}{\textbf{nDCG@20}} & \textbf{P@10} & \textbf{R@10}\\
\cmidrule(r){2-6} \cmidrule(lr){7-7} \cmidrule(lr){8-8}
& ND & LF & TN & LT & All & All & All \\
\midrule
\multicolumn{8}{c}{\textbf{without query augmentation}} \\
\midrule
\textbf{OpenAI v3 large}\\
formal corpus & 0.479 & 0.553 & 0.235 & 0.610 & 0.493 & 0.128 & 0.622\\
informal corpus & 0.591 & 0.527 & 0.365 & 0.451 & 0.512 & 0.134 & 0.628\\
formal+informal corpus & 0.607 & 0.549 & 0.267 & 0.622 & 0.545 & 0.140 & 0.677\\
\midrule
\textbf{$\text{E5}_{\text{mistral-7b}}$} \\
formal corpus & 0.576 & 0.633 & 0.294 & \textbf{0.774} & 0.593 & 0.132 & 0.687\\
informal corpus & 0.696 & 0.636 & 0.616 & 0.604 & 0.648 & 0.174 & 0.773\\
formal+informal corpus & \textbf{0.749} & 0.687 & 0.698 & 0.612 & 0.701 & 0.184 & 0.845\\
\midrule \midrule
\multicolumn{8}{c}{\textbf{with query augmentation}} \\
\midrule
\textbf{OpenAI v3 large}\\
formal corpus & 0.637 & 0.617 & 0.565 & 0.647 & 0.623 & 0.160 & 0.747\\
informal corpus & 0.603 & 0.650 & 0.810 & 0.575 & 0.640 & 0.166 & 0.783\\
formal+informal corpus & 0.681 & 0.657 & 0.772 & 0.703 & 0.691 & 0.178 & 0.837\\
\midrule
\textbf{$\text{E5}_{\text{mistral-7b}}$} \\
formal corpus & 0.688 & 0.685 & 0.631 & 0.694 & 0.680 & 0.178 & 0.847\\
informal corpus & 0.711 & 0.646 & 0.824 & 0.670 & 0.699 & 0.184 & 0.877\\
formal+informal corpus & 0.748 & \textbf{0.712} & \textbf{0.855} & 0.654 & \textbf{0.733} & \textbf{0.196} & \textbf{0.913}\\
\bottomrule
\end{tabular}
}
\caption{Results of ablation studies on document preparation and query augmentation. "All" indicates the performance averaged across the entire query set. In the query augmentation section, all queries are augmented to match the corpus type. Here, $\text{E5}_{\text{mistral-7b}}$ uses different task instructions for the document and query in the non-augmented query setting and the same task instructions for both in the query augmentation setting. Detailed task instructions can be found in Table \ref{table:Instructions_more}.}
\label{table:doc_query_ablation}
\end{table}

\begin{table}[h!]
\centering
\resizebox{\linewidth}{!}{
\begin{tabular}{
  |>{\centering\arraybackslash}m{2.7cm}
  |>{\centering\arraybackslash}m{1.2cm}
  |>{\raggedright\arraybackslash}m{8.6cm}|
}
\hline
\textbf{Query aug. \& Doc Type} & \textbf{Side} & \textbf{Task Instructions}\\ 
\hline
\multirow{2}{*}[-1.3em]{None \& Informal} & Query & \texttt{Given a math search query, retrieve theorems mathematically equivalent to the query}\\
\cline{2-3}
 & Doc & \texttt{Represent the given math theorem statement for retrieving related statement by natural language query}\\
\hline
\multirow{2}{*}[-2em]{None \& F+IF} & Query & \texttt{Given a math search query, retrieve theorems stated in bilingual Lean 4 + natural language that mathematically match the query}\\
\cline{2-3}
 & Doc & \texttt{Represent the given formal math statement written in Lean 4 concatenated with its natural language explanation for retrieving related statement by natural language query}\\
\hline
Formal \& Formal & Query \& Doc & \texttt{Retrieve math theorems stated in Lean 4 that are mathematically equivalent to the given one} \\
\hline
IF \& IF & Query \& Doc & \texttt{Retrieve math theorems that are mathematically equivalent to the given one} \\
\hline
\end{tabular}
}
\caption{Task Instructions for $\text{E5}_{\text{mistral-7b}}$ in ablation studies, supplementing Table \ref{table:Instructions_F_IF}. The abbreviation "aug." denotes augmentation, while "F" and "IF" represent formal and informal, respectively.
}
\label{table:Instructions_more}
\end{table}

\subsubsection{Ablation of Query Augmentations}\label{subsubsec:ablate_query}

The effect of query augmentation is evident when comparing the results in the upper and lower sections of Table \ref{table:doc_query_ablation}. As shown, augmented queries enhance the performance across all compared methods. Specifically, query augmentation significantly improves the ability of all methods to identify Theorem Names by expanding terms and providing richer context. Notably, the performance of $\text{E5}_{\text{mistral-7b}}$ on the formal + informal corpus without query augmentation is comparable to its performance on the same corpus with query augmentation. This makes it a cost-effective option for deploying our semantic search engine, as it eliminates the need to prompt an LLM with every query.

\section{Conclusion}\label{sec:conclusion}
In this paper, we introduce a semantic search engine designed to enable users to locate theorems in mathlib4 using informal queries. Specifically, we translate the formal statements of mathlib4 theorems into informal versions and develop our search engine to work with a corpus of informal-formal theorem pairs. Additionally, we construct a dataset to facilitate evaluation. Our comprehensive experiments on this dataset reveal that the best theorem retrieval performance is attained by augmenting the user's query appropriately and embedding the content of the corpus simultaneously. Consequently, our system employs a strategy that first augments the query, followed by a semantic search, thereby precisely aligning with the users' search intentions.

Our future research will focus on three primary directions. First, in terms of informalizing mathlib4, we aim to design guidelines for translation and provide examples of converting mathlib4 statements to informal language for LLMs, enhancing the informal corpus's quality. Second, regarding the mathlib4 semantic search benchmark, we plan to continually enlarge the query set, aiming for a more comprehensive benchmark. Lastly, for the semantic search engine itself, we intend to fine-tune a text embedding model on the task of theorem retrieval, aiming to improve the search engine's performance.

\section{Limitations}
When translating mathlib4 theorems into their informal versions, we extract the related definitions through the hyperlinks in the theorem statements. This approach to extracting dependencies can sometimes be inaccurate and may omit certain dependencies. A more precise method is to extract the dependencies of a theorem by interacting with Lean's language server.

Another limitation of our work is the use of $\text{E5}_{\text{mistral-7b}}$, a 7B embedding model, in our search engine. This model may pose challenges when deploying the search engine on resource-constrained devices like laptops. In future work, we aim to improve the performance of smaller embedding models for theorem retrieval tasks to enable broader accessibility.

\section*{Acknowledgment}
This work is supported in part by the New Cornerstone Investigator Program.

\bibliography{ref}

\appendix

\section{Example of Informalized Mathlib4 Statements}
\label{sec:example}
In this section, we provide an example comparing the informalized mathlib4 statement generated by \citep{DBLP:journals/corr/abs-2311-03755}, which used GPT-4 without any additional context, to the statement produced by our approach that incorporates additional context.
\subparagraph*{Formal Statement:} theorem dite\_eq\_iff \{$\alpha$ : Sort u\_2\} \{P : Prop\} $[\text{Decidable P}]$ \{c : $\alpha$\} \{A : P → $\alpha$\} \{B : $\lnot$P → $\alpha$\} : dite P A B = c $\leftrightarrow$ ($\exists$ h, A h = c) $\lor$ $\exists$ h, B h = c
\subparagraph*{Informal Statement \citep{DBLP:journals/corr/abs-2311-03755}\footnote{\href{https://github.com/albertqjiang/MMA/tree/main/data}{https://github.com/albertqjiang/MMA/tree/main/data}}:} dite P A B equals c if and only if there exists a proof h such that A h equals c, or there exists a proof h such that B h equals c.

\subparagraph*{Informal Statement (ours):} Dependent if-then-else equivalence: For any proposition P that is decidable, and any elements c, A, and B, the expression `if P then A else B' is equal to c if and only if either there exists a proof h such that A h is equal to c, or there exists a proof h such that B h is equal to c.

In this example, the informalized statement provided by \citep{DBLP:journals/corr/abs-2311-03755} retains the "dite" notation, which may be difficult for readers unfamiliar with mathlib4 naming conventions. In contrast, our informalized statement translates "dite" into "dependent if-then-else," making the statement more accessible. This translation was achieved by extracting the definition of "dite" and its documentation string "Dependent if-then-else..." in the prompt, demonstrating how additional context enhances the interpretation and conversion of formal theorems into informal language.

\section{Prompt for Query Augmentation}
\label{sec:append_qa_prompt}
\begin{figure}[hbt!]
\centering
\includegraphics[width=\columnwidth]{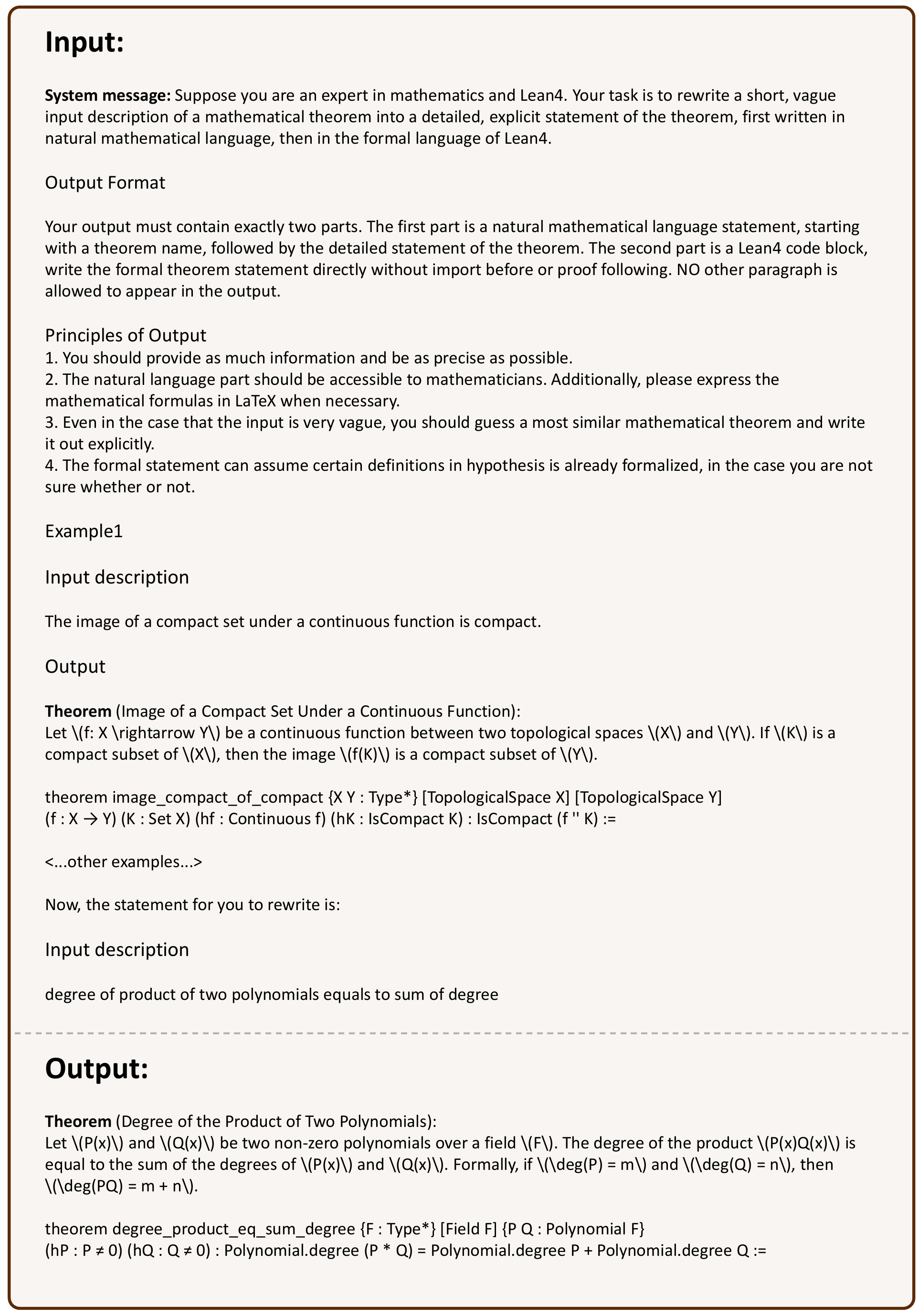}
\caption{Prompt for query augmentations.}
\label{fig:qa_prompt}
\end{figure}

In this section, we present the prompt template for query augmentation, as illustrated in Figure \ref{fig:qa_prompt}. We outline specific principles for query augmentation, highlighting the importance of providing precise information, using LaTeX for mathematical expressions, and clarifying ambiguous inputs. Additionally, we offer examples of query augmentations to enhance understanding of the task.

\section{Implementation Details}
\label{sec:append_detail}
In this section, we provide the implementation details of our experiments.

We use default task instructions in UAE-Large-V1. For $\text{E5}_{\text{mistral-7b}}$, we employ different task instructions for the document and query in the non-augmented query setting and the same task instructions for both in the query augmentation setting, as shown in Table \ref{table:Instructions_F_IF}. Notably, the original implementation of $\text{E5}_{\text{mistral-7b}}$ omits document-side task instructions to reduce computational costs during document indexing in multiple retrieval tasks. Given the complexity of the theorem retrieval task, we have modified this approach to use two-sided prompts.

During evaluation, inputs to the embedding model exceeding 4096 characters were truncated due to GPU memory limitations. For all tested models, the corpus embedding process was completed within three hours on a single Nvidia V100 GPU with 16GB of GPU memory.
\begin{table}[h!]
\centering
\resizebox{\linewidth}{!}{
\begin{tabular}{
  |>{\centering\arraybackslash}m{2.5cm}
  |>{\centering\arraybackslash}m{1.2cm}
  |>{\raggedright\arraybackslash}m{8.8cm}|
}
\hline
\textbf{Query aug. \& Doc Type} & \textbf{Side} & \textbf{Input with Task Instructions}\\ 
\hline
\multirow{2}{*}[-2em]{None \& Formal} & Query & \texttt{"Instruct: Given a math search query, retrieve theorems stated in Lean 4 that mathematically match the query \textbackslash n Query:\{F+IF augmented query\}"}\\
\cline{2-3}
 & Doc & \texttt{"Instruct: Represent the given formal math statement written in Lean 4 for retrieving related statement by natural language query \textbackslash n Doc:\{Formal statement\}"}\\
\hline
\multirow{2}{*}[-2em]{F+IF \& F+IF} & Query & \texttt{"Instruct: Retrieve math theorems stated in bilingual Lean 4 + natural language that are mathematically equivalent to the given one \textbackslash n Query:\{query\}"}\\
\cline{2-3}
 & Doc & \texttt{"Instruct: Retrieve math theorems stated in bilingual Lean 4 + natural language that are mathematically equivalent to the given one \textbackslash n Doc:\{F + IF statement\}"}\\
\hline
\end{tabular}
}
\caption{Task instructions used in $\text{E5}_{\text{mistral-7b}}$. Here "aug." stands for augmentation, and F and IF stands for formal and informal respectively. The "None \& Formal" setting is used as baseline, and "F+IF \& F+IF" is used in our method.}
\label{table:Instructions_F_IF}
\end{table}

\section{Ablation Study on Task Instructions}
\label{sec:append_abl}
In this section, we investigate the impact of task instructions on the theorem retrieval performance. We present a visualization of the embeddings of all 50 queries, generated by four embedding engines in our benchmark, using t-SNE\citep{JMLR:v9:vandermaaten08a}. We use the default task instruction for UAE-Large-V1. For $\text{E5}_{\text{mistral-7b}}$, we test three instructions: an empty instruction, a mathematics retrieval instruction ("Given a math search query, retrieve theorems mathematically equivalent to the query"), and a Lean 4 retrieval instruction ("Given a math search query, retrieve Lean 4 written theorems that mathematically match the query").

\begin{figure}[h!]
\centering
\begin{subfigure}{0.46\columnwidth}
    \centering
    \includegraphics[width=\textwidth]{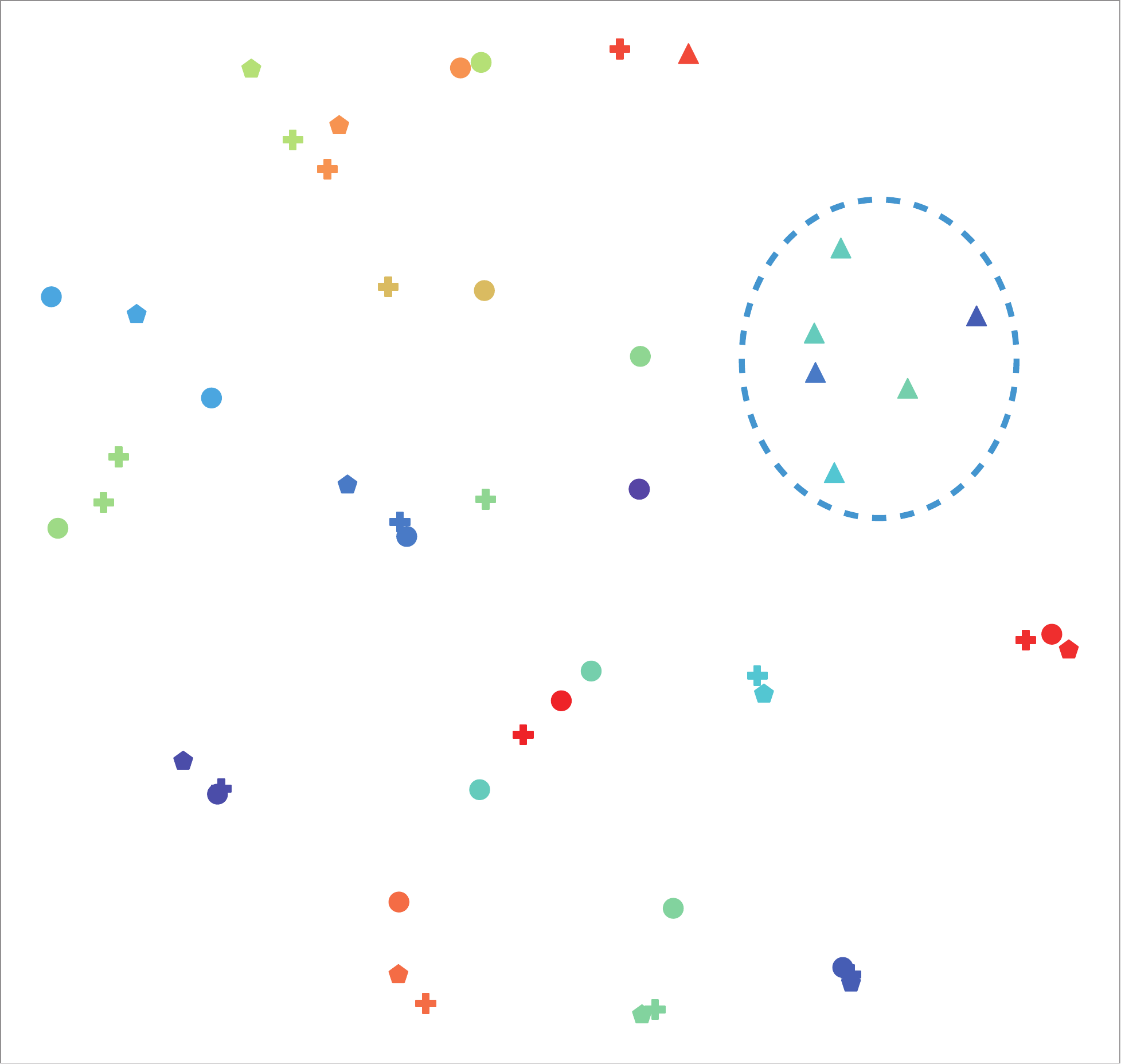}
    \caption{text-embedding-ada-002 (without task instruction)}
    \label{fig:Ada002}
\end{subfigure}
\hfill
\begin{subfigure}{0.46\columnwidth}
    \centering
    \includegraphics[width=\textwidth]{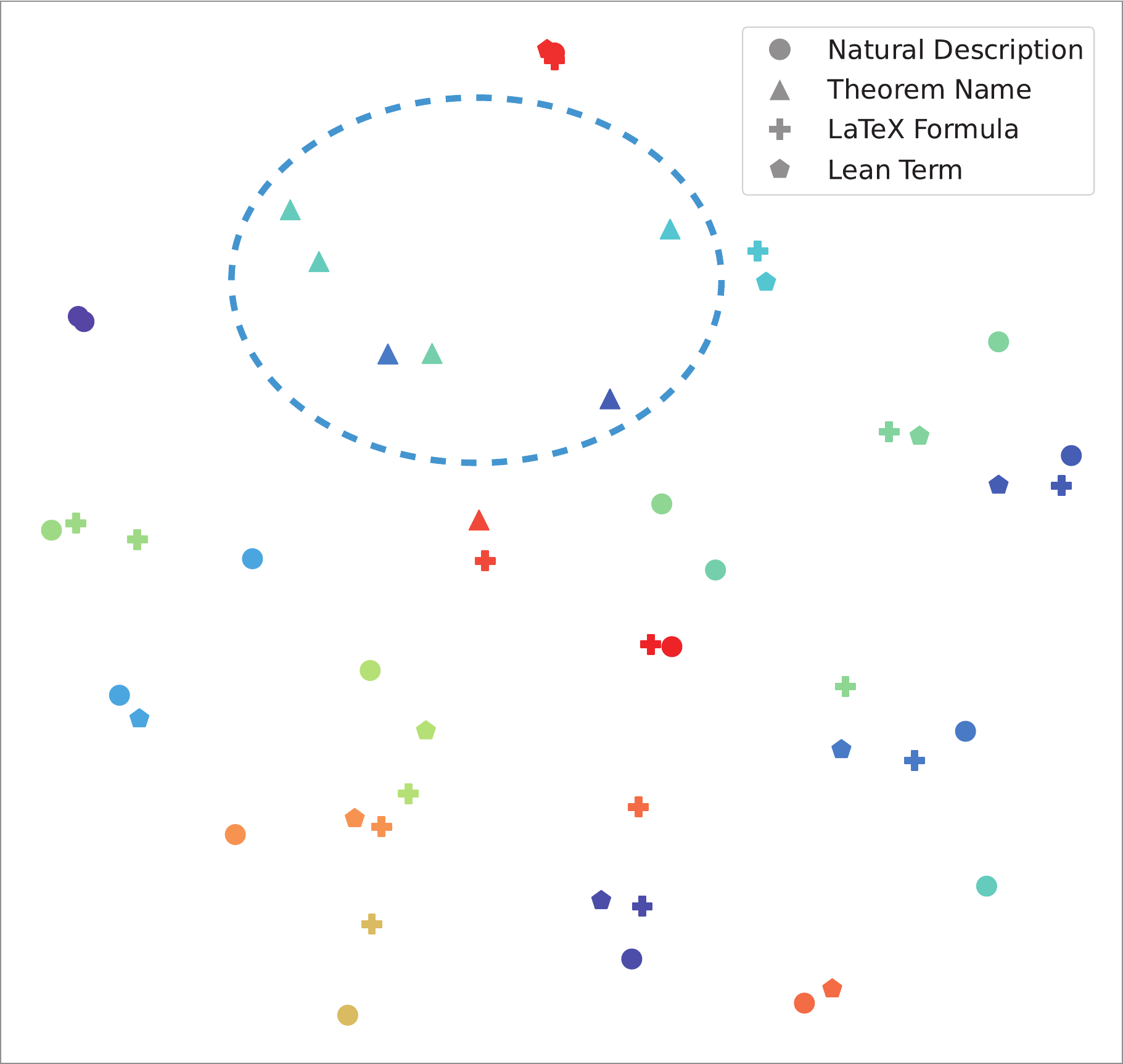}
    \caption{UAE-Large-V1 with default instruction}
    \label{fig:UAE}
\end{subfigure}
\hfill
\begin{subfigure}{0.46\columnwidth}
    \centering
    \includegraphics[width=\textwidth]{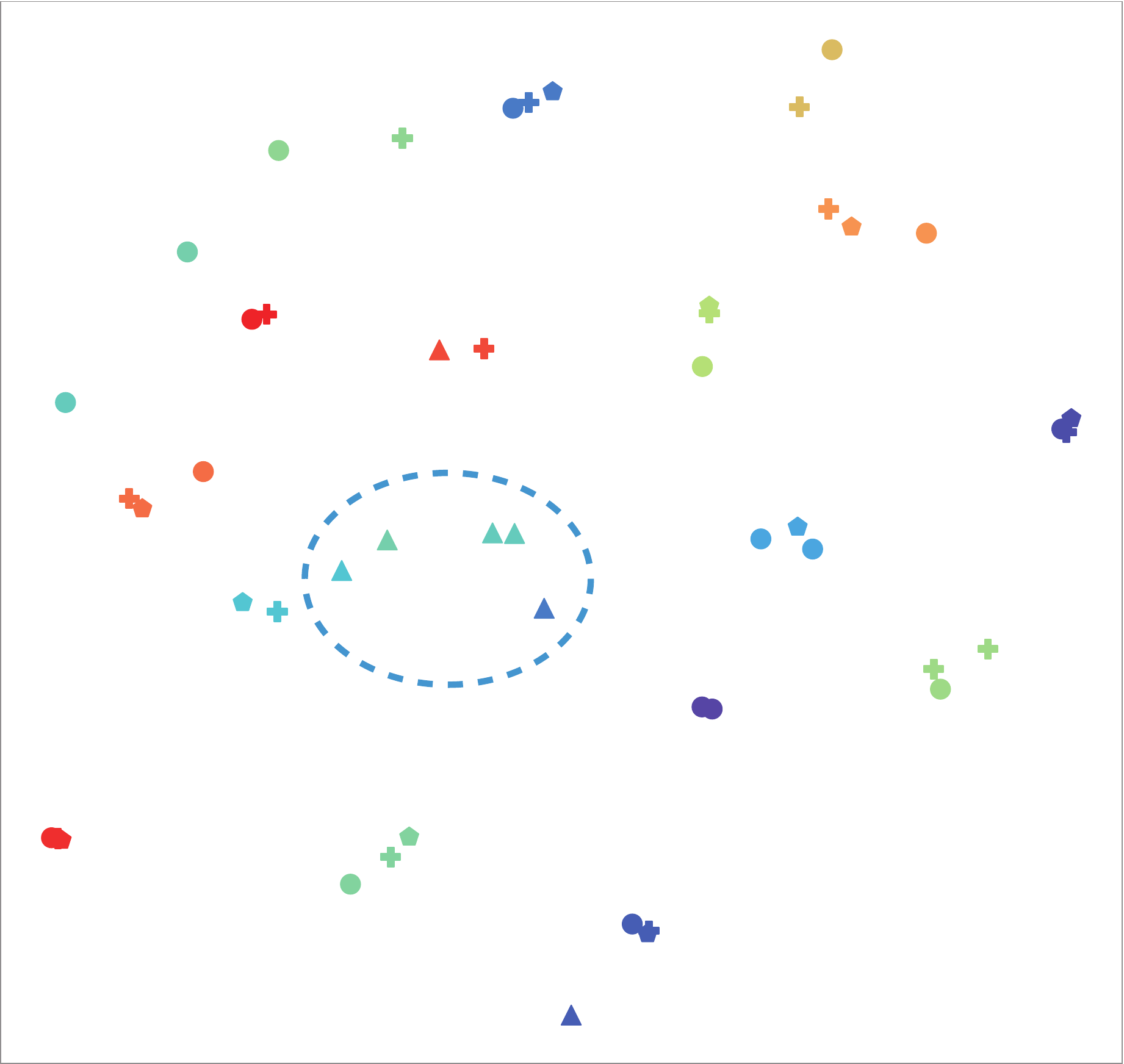}
    \caption{text-embedding-3-large (without task instruction)}
    \label{fig:Ada003}
\end{subfigure}
\hfill
\begin{subfigure}{0.46\columnwidth}
    \centering
    \includegraphics[width=\textwidth]{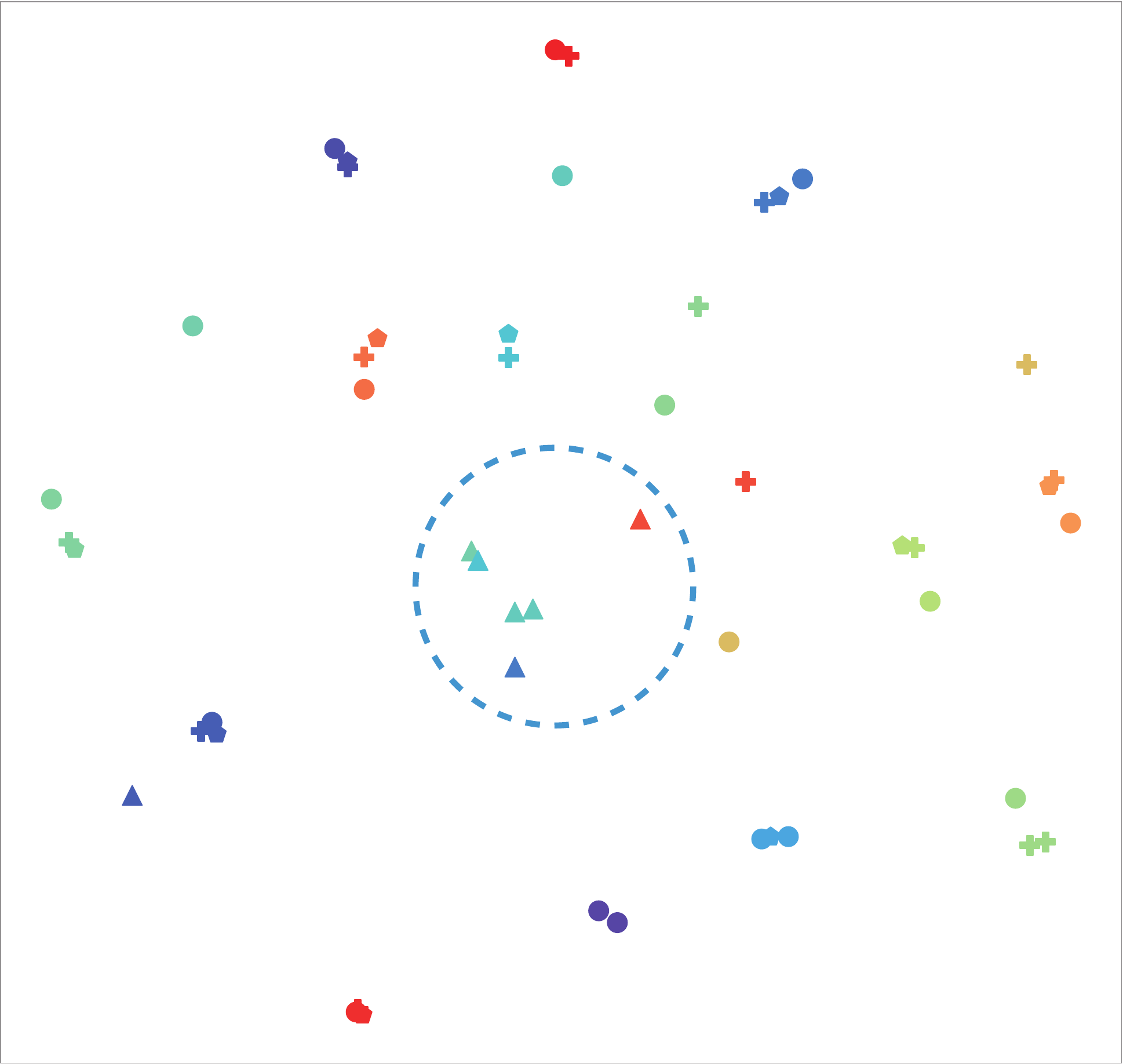}
    \caption{$\text{E5}_{\text{mistral-7b}}$ with empty instruction}
    \label{fig:MIS_empty}
\end{subfigure}
\hfill
\begin{subfigure}{0.46\columnwidth}
    \centering
    \includegraphics[width=\textwidth]{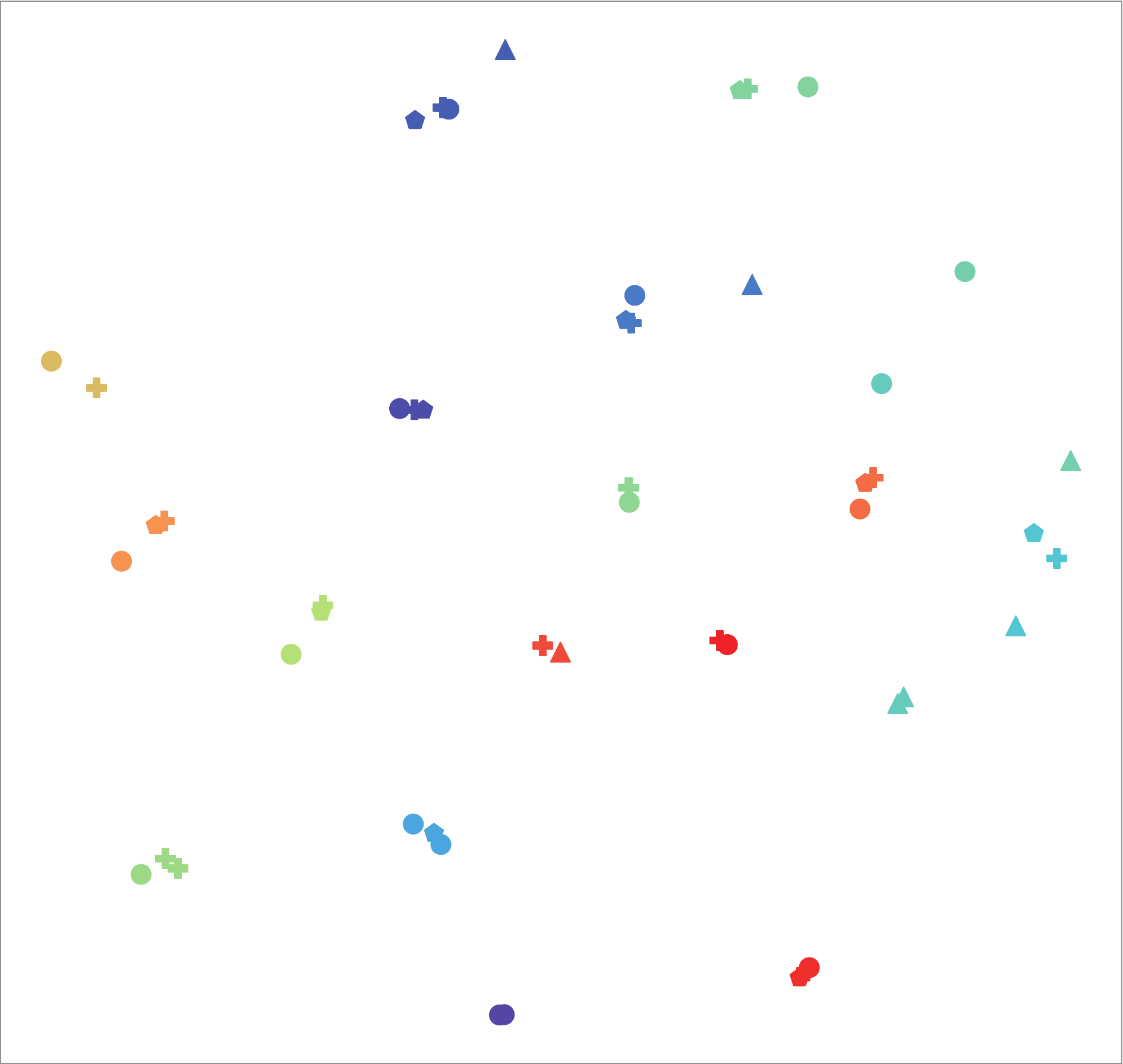}
    \caption{$\text{E5}_{\text{mistral-7b}}$ with Math retrieving instruction}
    \label{fig:MIS_math}
\end{subfigure}
\hfill
\begin{subfigure}{0.46\columnwidth}
    \centering
    \includegraphics[width=\textwidth]{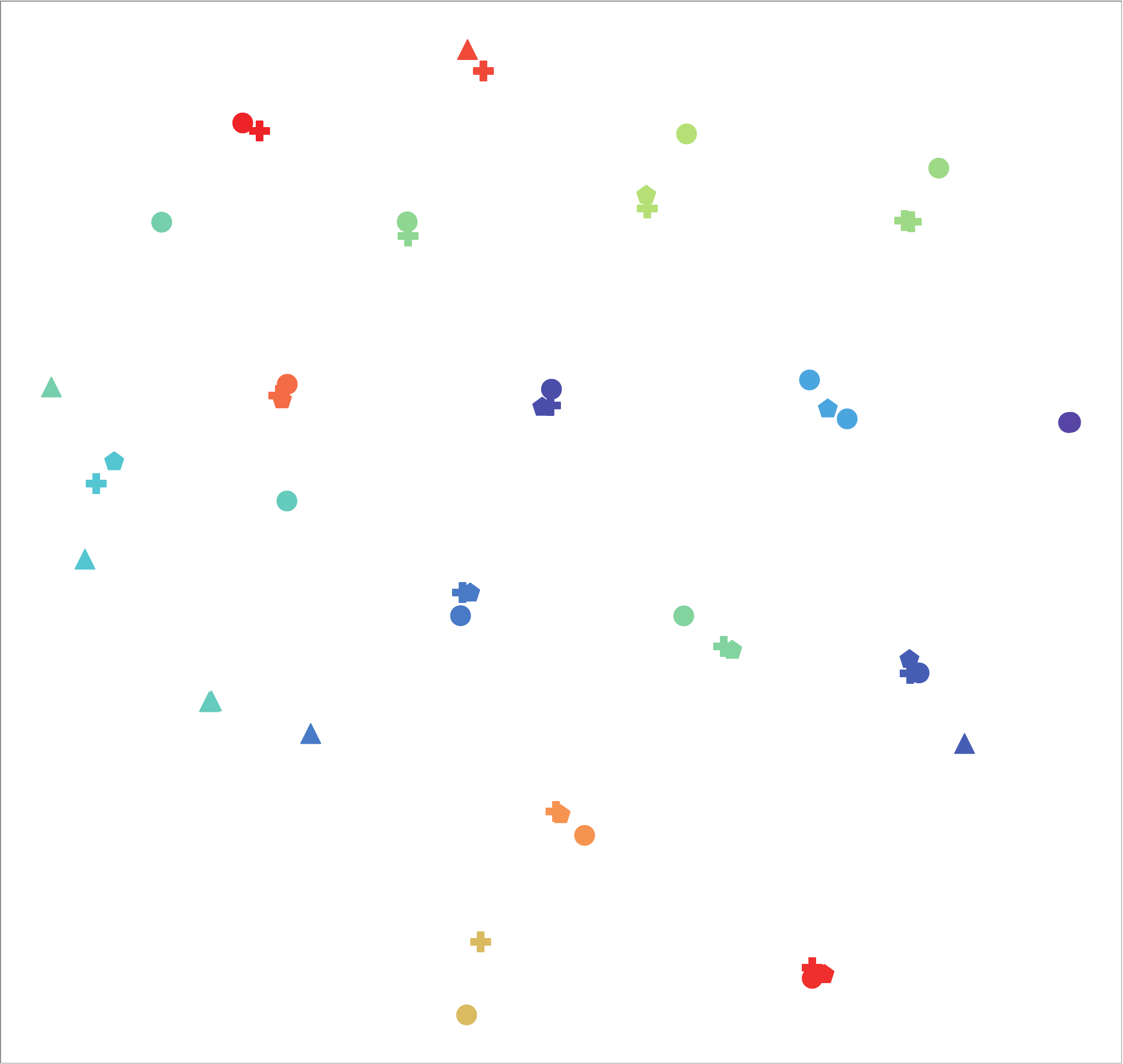}
    \caption{$\text{E5}_{\text{mistral-7b}}$ with Lean 4 retrieving instruction}
    \label{fig:MIS_Lean4}
\end{subfigure}
\caption{t-SNE visualization of the embeddings of all queries in our benchmark. Each dot denotes a distinct query, with queries within the same query group sharing identical colors. The shape of the markers differentiates the four distinct query categories. Theorem name clusters are emphasized with circles.}
\label{fig:t-SNE}
\end{figure}

As illustrated in Figure \ref{fig:t-SNE}, clusters of theorem names appear in the results produced by all four embedding models. However, these clusters vanish when we provide $\text{E5}_{\text{mistral-7b}}$ with mathematics-aware task instructions. Given that we assume all queries within the same query group (denoted by identical colors in Figure \ref{fig:t-SNE}) have the same search intent, they should be proximal on the graph, as t-SNE maintains the relative distance relationships between vectors. Thus, the more effective the embedding engine, the more closely the dots of the same color group together. A comparison of subfigures \ref{fig:Ada003} and \ref{fig:MIS_empty} with \ref{fig:Ada002} and \ref{fig:UAE} reveals that text-embedding-3-large and $\text{E5}_{\text{mistral-7b}}$ demonstrate superior performance in discerning mathematical search intentions, aligning with our main results (Table \ref{table:main_results}).

The presence of theorem name clusters in these subfigures, however, contradicts this principle, suggesting that the embedding models perceive these theorem names as more similar to each other rather than correctly associating them with their respective query groups. While this might be acceptable for other retrieval tasks, it is inappropriate for our mathematical information retrieval context. In contrast, the absence of theorem clusters in subfigures \ref{fig:MIS_math} and \ref{fig:MIS_Lean4} suggests that these embedding methods successfully group theorem names with their corresponding statements, despite their notable differences in appearance and structure. This indicates their proficiency in comprehending mathematical theorems and recognizing search intents. In summary, appropriate task instructions enhance the embedding models' sensitivity to the search intents of mathematical concepts, thereby improving retrieval effectiveness on mathlib4.

\section{Models for Query Augmentation}
In this section, we compare the performance of two LLMs, DeepSeek-V2-Chat and GPT-4, for query augmentation. Table \ref{table:qa_model} presents the performance of both models using the informal+formal corpus setting, with results averaged over 10 runs.

\begin{table}[h!]
\centering
\resizebox{\linewidth}{!}{
\begin{tabular}{lccccc}
\toprule
\textbf{API}  & \textbf{nDCG@20} & \textbf{P@10} & \textbf{R@10} & \textbf{Cost per 1M input tokens}	   &  \textbf{Cost per 1M output tokens} \\
\midrule
\textbf{gpt-4-0125-preview	} & 0.721 & 0.195 & 0.911 & \$10.00 & \$30.00\\
\textbf{deepseek-chat} & 0.749  &  0.202 & 0.933 & \$0.14 & \$0.14 \\
\bottomrule
\end{tabular}
}
\caption{Performance of DeepSeek-V2-Chat and GPT-4 for query augmentation.}
\label{table:qa_model}
\end{table}

As shown, DeepSeek-V2-Chat not only outperforms GPT-4 in terms of effectiveness but also offers a significantly lower cost. The results suggest that DeepSeek-V2-Chat is a more efficient and cost-effective choice for query augmentation tasks.

\end{document}